\documentclass{article}
\usepackage{graphicx}
\usepackage{hyperref}
\usepackage{authblk}

\begin{document}
\title{Three dimensional L\'evy HBT results from PHENIX%
\thanks{Presented by M. Csan\'ad at XIII Workshop on Particle Correlations and Femtoscopy, 22-26 May 2018, The Henryk Niewodnicza\'nski Institute of Nuclear Physics PAN, Krak\'ow, Poland}}%

\author[1]{B\'alint Kurgyis for the PHENIX Collaboration}
\affil[1]{E\"otv\"os Lor\'and University, Hungary\\ H-1117 Budapest, P\'azm\'any P. s. 1/A}

\maketitle
\begin{abstract}
In heavy-ion collisions we measure the Bose-Einstein correlation functions to map out the femtoscopic homogeneity region of the particle emitting source. Previously, a one dimensional measurement with a L\'evy shaped source was done at the PHENIX experiment.  In this paper, we present the three dimensional L\'evy HBT results of PHENIX and their comparison with the one dimensional results.
\end{abstract}
  
\section{Introduction}
The PHENIX experiment at the Relativistic Heavy Ion Collider (RHIC) has collected a large amount of data with different collision systems (p+p, p+A, A+A) at different collision energies \cite{Adcox:2004mh}. This gives the chance to thoroughly investigate the created quark-gluon plasma (QGP) and its properties. The QGP cools down and from it hadrons are created. We can map out the femtoscopic homogeneity region of this particle emitting source by measuring the Bose-Einstein correlation functions of identical bosons. The shape of the source was often assumed to be a Gaussian \cite{Adler:2004rq,Adamczyk:2014mxp}. However, there were indications of non-Gaussian behavior \cite{Adler:2006as,Afanasiev:2007kk,Csanad:2005nr}. More recently PHENIX measured the Bose-Einstein correlation functions in one dimension \cite{Adare:2017vig} and found that the data can be best described by a more general, L\'evy shaped source. The L\'evy parameters were only measured in one dimension so far, but it is important to investigate the appropriacy of the L\'evy type source assumption in three dimensional HBT measurements. Here we present the preliminary results of a three dimensional L\'evy analysis of same charged pions from $\sqrt{s_{_\mathrm{NN}}}=200$ GeV Au+Au data.
\section{The L\'evy type source}
One possible explanation for the appearance of L\'evy distributions in heavy-ion collisions is the rapidly expanding medium, where the mean free path increases, thus anomalous diffusion emerges \cite{Csanad:2007fr}. The L\'evy distribution is characterized by the L\'evy exponent $\alpha$. At two special values of $\alpha$ the L\'evy distribution can be formulated analytically: $\alpha=2$ leads to a Gaussian distribution and $\alpha=1$ leads to a Cauchy distribution. In general the L\'evy stable distribution can be expressed with an integral formula:
\begin{equation}
\mathcal{L}(\alpha,\mathbf{R},\mathbf{r})=\frac{1}{(2\pi)^{3}} \int \mathrm{d}^3q e^{i\mathbf{qr}} e^{-\frac{1}{2}|\mathbf{qR}|^{\alpha}}.
\end{equation}
We can measure the three dimensional L\'evy HBT scale parameters ($R_\mathrm{out}$, $R_\mathrm{side}$, $R_\mathrm{long}$) by comparing data to Bose-Einstein correlation functions of L\'evy sources. These scale parameters differ from the usually measured Gaussian radii but still they describe the homogeneity region of the particle emitting source.

Furthermore, it is important to consider the role of pions coming from resonance decays in the system. We can take these into account via the core-halo model \cite{Csorgo:1994in,Bolz:1992hc}, where we assume a two component source. One is a hot, dense and hydrodynamically expanding core with a characteristic size $\sim 10$ $\mathrm{ fm}/c$. The core emits the primordial pions and yields the correlation function that we can measure. The other component is a rather wide spatial distribution of the decay products of long lived resonances with a characteristic scale much larger than $10$ $\mathrm{ fm}/c$, which transforms into a very narrow peak in the relative momentum space, thus it is experimentally unresolvable. It turns out that $\lambda$ correlation strength parameter (intercept of the correlation function at $\mathbf{q}=0$) can be expressed as: $\lambda=N^2_\mathrm{core}/(N_\mathrm{core}+N_\mathrm{halo})^2$, where $N_\mathrm{core}$ and $N_\mathrm{halo}$ are the number of pions coming from the core and the halo, respectively. Hence, this way we can describe the ratio of primordial pions in the system by measuring lambda.

With the above assumptions we end up with the following form for the interaction free two-particle correlation function \cite{Csorgo:2003uv}:
\begin{equation}
C^{(0)}_2(\mathbf{q};\mathbf{R},\lambda,\alpha)=1+\lambda\exp\left(-|\mathbf{qR}|^\alpha\right).
\end{equation}
Additionally, one needs to consider the Coulomb interaction between the charged pions. We can take this into account in the following way:
\begin{equation}
C^{(C)}_2(\mathbf{q};\mathbf{R},\lambda,\alpha)=K(\mathbf{q};\mathbf{R},\lambda,\alpha)\cdot C^{(0)}_2(\mathbf{q};\mathbf{R},\lambda,\alpha),
\end{equation}
where the $K(\mathbf{q})$ is the Coulomb correction factor. In this analysis we used a very detailed numerical table for $K(\mathbf{q})$ and fitted the data with an iterative method as described in ref.~\cite{Adare:2017vig}. Here the assumption of a spherically symmetric source is used.
\section{Three dimensional L\'evy HBT at PHENIX}
\begin{figure}[]
\centerline{%
\includegraphics[width=12.5cm]{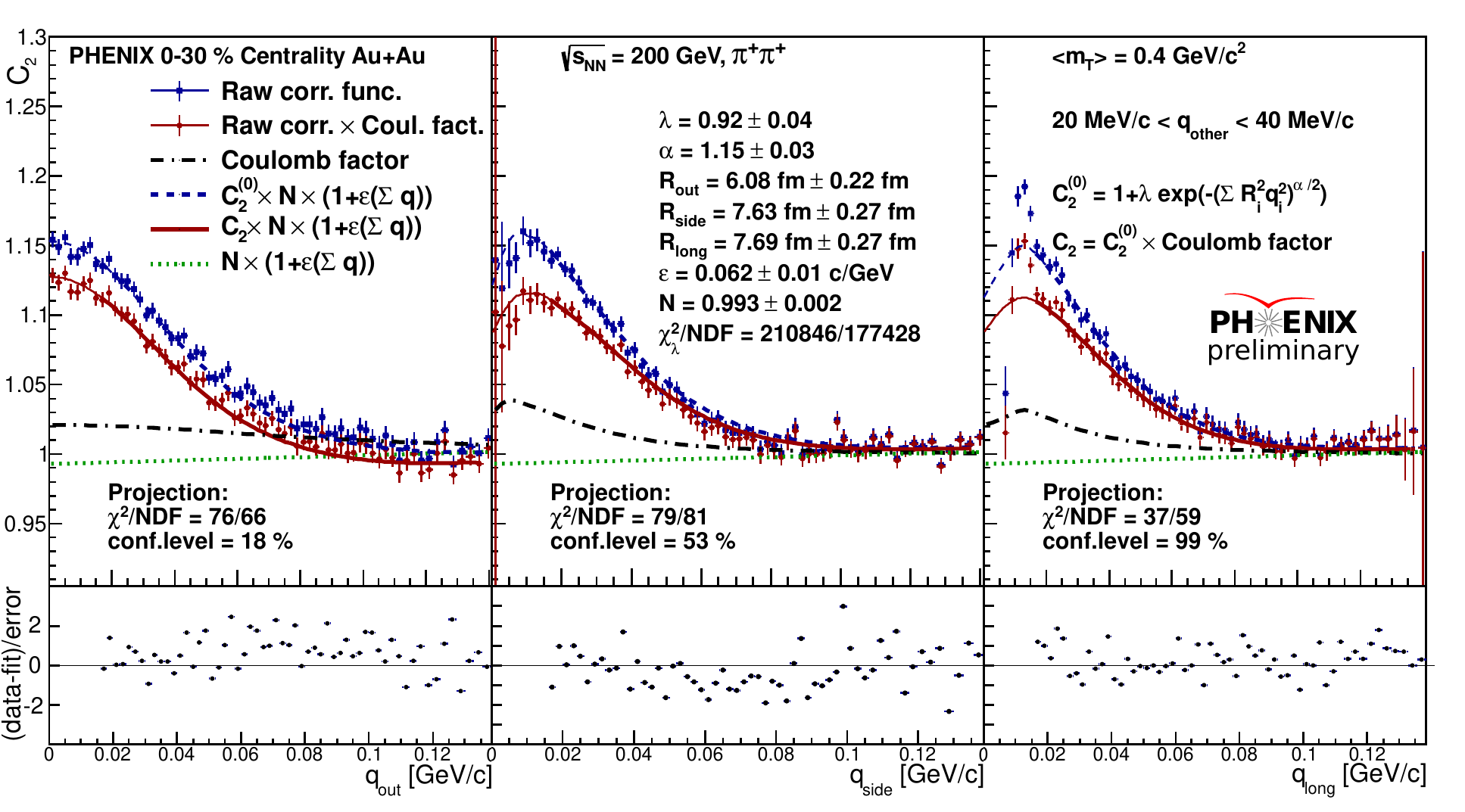}}
\caption{An example plot of the correlation functions measured in three dimensions by showing the projections along the out, side and long directions. We can see the raw data points and the fitted theoretical curve. The first few points are not fitted because of two track resolution. Also, the the Coulomb corrected correlation functions are shown. }
\label{fig:c_proj}
\end{figure}

We analyzed $\sqrt{s_{_\mathrm{NN}}}=200$ GeV Au+Au data from the 2010 data taking period at the PHENIX experiment. The two-particle relative momentum correlation functions were measured for $\pi^+\pi^+$ and $\pi^-\pi^-$ pairs from the $0-30\%$ centrality events in 31 different transverse momentum ($m_\mathrm{T}$) bins. For the measurements we used Bertsch-Pratt coordinates (out, side, long), in the longitudinally comoving system. Then, we fitted the data with correlation functions from a L\'evy shaped source. This way we could extract the transverse momentum dependence of the model parameters. For the fitting we used a modified Poisson likelihood term ($\chi^2_\lambda$) as in ref.~\cite{Ahle:2002mi} that considers the Poisson error for the bin counts. Fig.~\ref{fig:c_proj} shows an example of the projections of the measured three dimensional correlation function along the out, side and long directions and also the projected fit.

\begin{figure}[htb]
\centerline{%
\includegraphics[width=12.5cm]{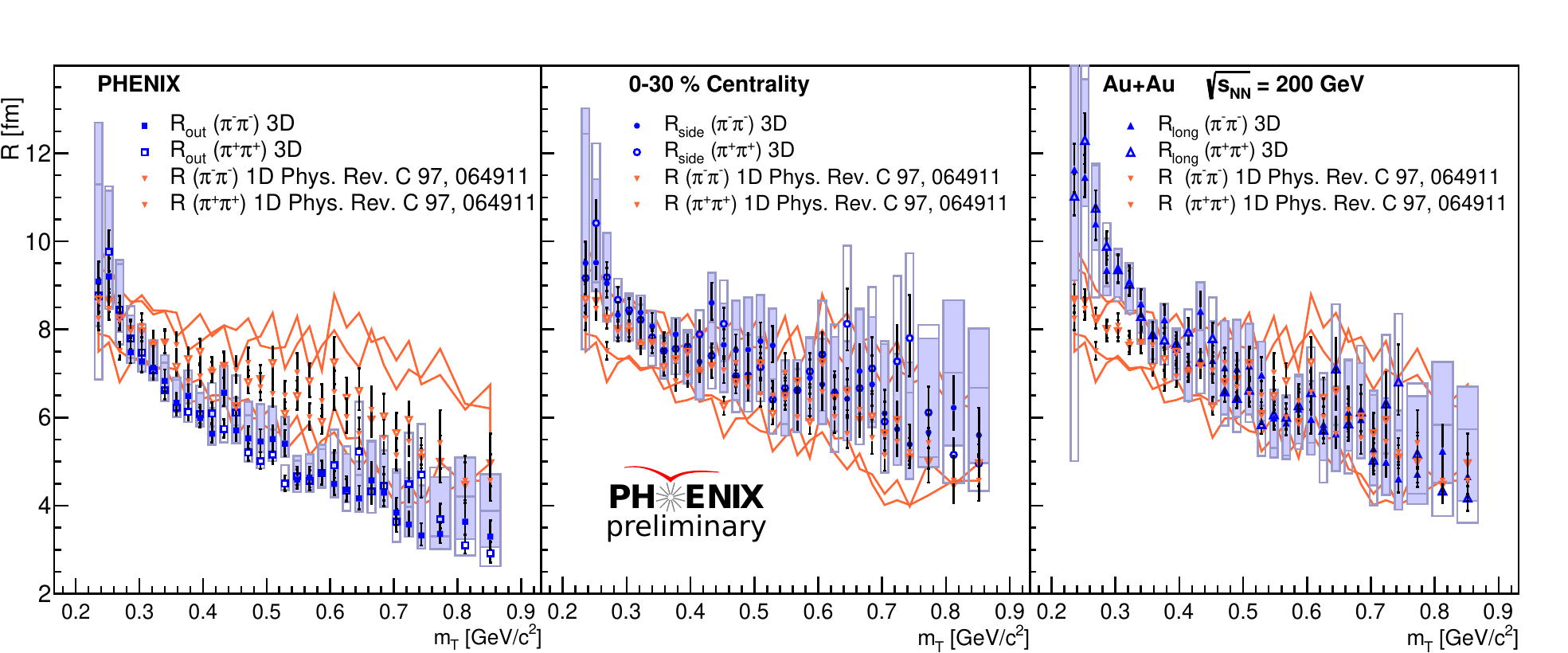}}

\centerline{
\includegraphics[width=6.4cm]{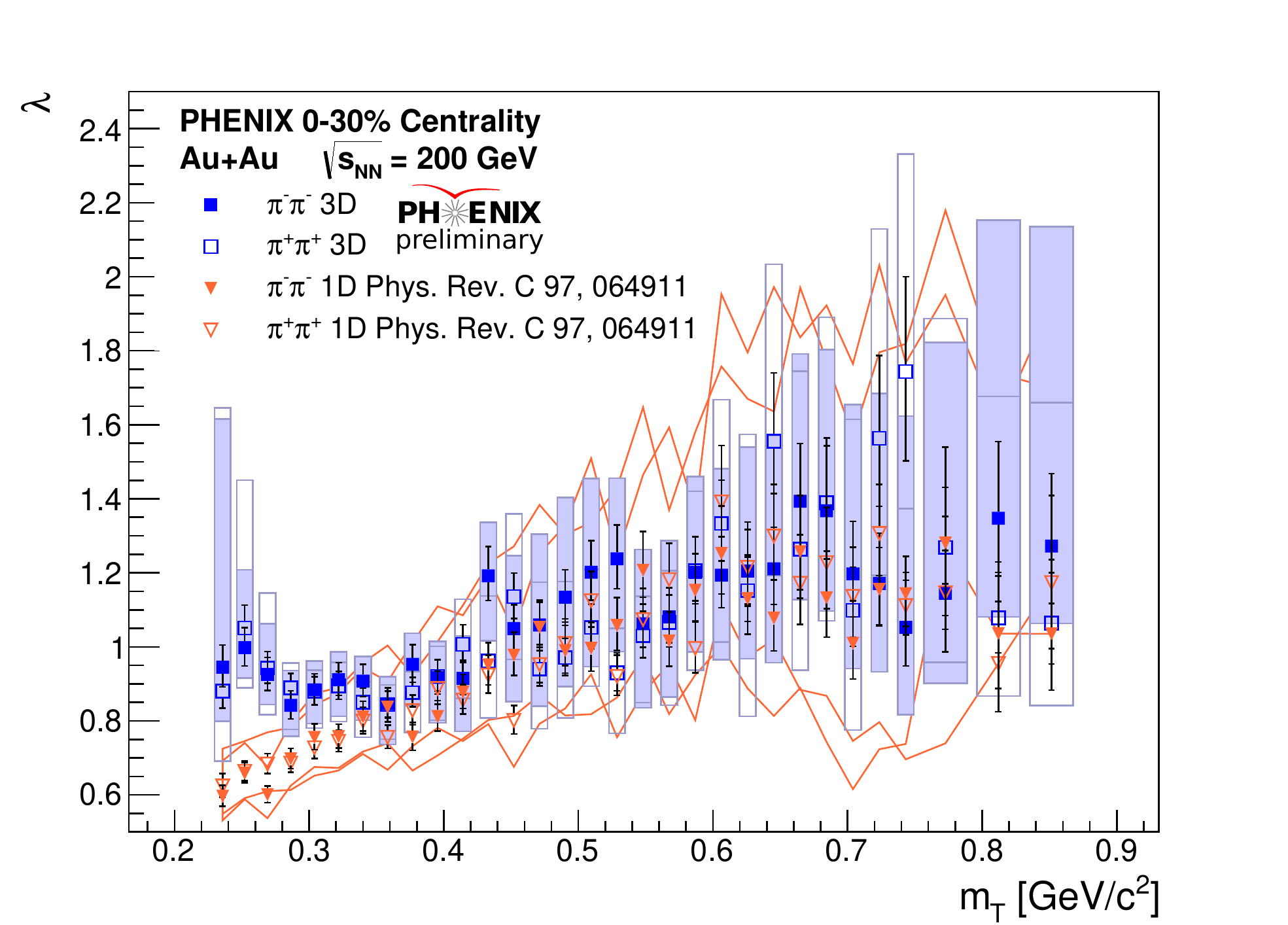}\includegraphics[width=6.4cm]{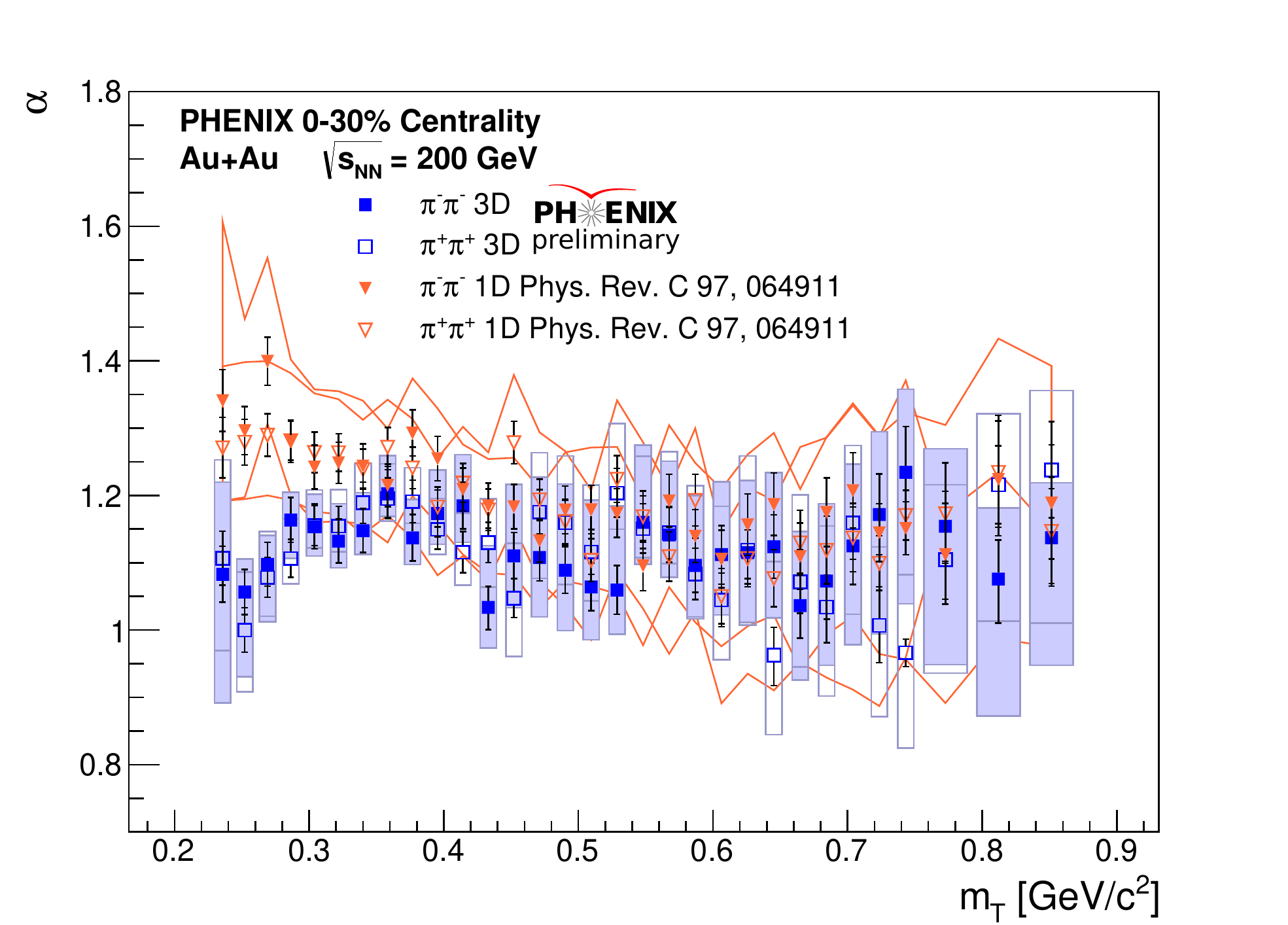}}
\caption{On the top the transverse momentum dependence of the $R_\mathrm{out}$, $R_\mathrm{side}$, $R_\mathrm{long}$ L\'evy scale parameters from this measurement and the previous one dimensional results are also shown. At the bottom we can see the results of this measurement compared with the previous one dimensional results as the function of transverse mass: the L\'evy exponent ($\alpha$) on the right and the strength of the correlation ($\lambda$) on the left.}
\label{fig:pars}
\end{figure}

In addition to the investigation of the transverse momentum dependence of the L\'evy parameters we also did a comparison with previous one dimensional results from L\'evy HBT at PHENIX \cite{Adare:2017vig}. Fig.~\ref{fig:pars} shows the transverse momentum dependence of the L\'evy model parameters. On the top panel we can see the L\'evy scale parameters ($R_\mathrm{out}$, $R_\mathrm{side}$, $R_\mathrm{long}$) as a function of $m_\mathrm{T}$. At the first few $m_\mathrm{T}$ bins the source is rather asymmetric, which might affect the accuracy of the way we handled the Coulomb interaction. Except the first few points the radii are consistent with the one dimensional radius. The often measured Gaussian radii usually show a hydrodynamical scaling ($R\propto1/\sqrt{m_\mathrm{T}}$) across various collision centralities and particle types \cite{Adler:2004rq,Afanasiev:2009ii}. Here, we are not measuring the Gaussian radii, but the L\'evy scale parameters and these still follow the hydrodynamical scaling. The the bottom left plot of fig. \ref{fig:pars} shows the transverse momentum dependence of the correlation strength ($\lambda$). Apart from a small deviation at the lower $m_\mathrm{T}$ bins the new results are consistent with the published one dimensional PHENIX results \cite{Adare:2017vig}. In the bottom right plot of fig.~\ref{fig:pars} we can see the L\'evy exponent ($\alpha$) as a function of the transverse momentum measured in this analysis and also previously in ref.~\cite{Adare:2017vig}. The results are consistent with each other, the only deviation is at the low $m_\mathrm{T}$ region. This might occur because the source is the most asymmetric in that region, therefore a non-spherically symmetric Coulomb correction may have to be used. These plots indicate that the best description of the data is given by the general L\'evy shaped source (with $\alpha \sim 1.2$) and not by a Gaussian ($\alpha=2$) or Cauchy ($\alpha=1$) source.

\section{Summary}
We measured the three dimensional Bose-Einstein correlation functions and investigated the transverse mass dependence of L\'evy parameters. We found that the L\'evy exponent and the strength of the correlation are compatible with the one dimensional measurement. This means, that the best description of the data is given by a general L\'evy shaped source with an exponent $\alpha \sim1.2$, which is far from the Gaussian case ($\alpha=2$) and also differs from the Cauchy distribution ($\alpha=1$). The transverse mass dependence of the L\'evy scale parameters ($R_\mathrm{out}$, $R_\mathrm{side}$, $R_\mathrm{long}$) shows us that we have an asymmetric source, especially at the low $m_\mathrm{T}$ region. Additionally, we observed the hydrodynamical scaling ($R\propto1/\sqrt{m_\mathrm{T}}$) of the homogeneity length'. For the approximative Coulomb correction a more advanced calculation is needed, to treat the asymmetric source appropriately. With this we hope to understand the low $m_\mathrm{T}$ behavior of the parameters better, where the preliminary results indicate the most asymmetric source.

\section*{Acknowledgements}
The author expresses gratitude for the support of Hungarian NKIFH grant No.~FK-123842. B. Kurgyis was supported by the \'UNKP-18-1-I-ELTE-320 New National Excellence Program of the Hungarian Ministry of Human Capacities.

\bibliographystyle{unsrt}

\end{document}